\documentstyle{aipproc}
\begin{document}
\title{ Nonlinear Dynamics of Accelerator via Wavelet Approach}
\author{A.N.~Fedorova and  M.G.~Zeitlin}
\address{Institute of Problems of Mechanical Engineering,\\
 Russian Academy of Sciences, Russia, 199178, St.~Petersburg,\\
  V.O., Bolshoj pr., 61, e-mail: zeitlin@math.ipme.ru}
\maketitle
\thispagestyle{empty}
\begin{abstract}
In this paper we
present the applications of methods from
wavelet analysis to polynomial approximations for
a number of accelerator physics problems.
In the general case we have the solution as
a multiresolution expansion in the base of compactly
supported wavelet basis. The solution is parametrized by
the solutions of two reduced algebraical problems, one is
nonlinear and the second is some linear
problem, which is obtained from one of the next wavelet
constructions: Fast Wavelet Transform, Stationary
Subdivision Schemes, the method of Connection
Coefficients. According to the orbit method and by using construction 
from the geometric quantization theory we construct the symplectic 
and Poisson structures associated with generalized wavelets by
using metaplectic structure. We consider wavelet approach to
the calculations of Melnikov functions in the theory of homoclinic
chaos in perturbed Hamiltonian systems and for parametrization of
Arnold--Weinstein curves in Floer variational approach.
\end{abstract}
\section*{ Introduction.}
In this paper we consider the following problems: the calculation
of orbital motion in storage rings, some aspects of symplectic invariant
 approach to wavelet computations, Melnikov functions approach
in the theory of homoclinic chaos, the
calculation of Arnold-Weinstein curves (periodic loops) in Hamiltonian
systems.
The key point in
the solution of these problems is the use of the methods of
wavelet analysis,
relatively novel set of mathematical methods, which gives
us a possibility to work with well-localized bases in functional
spaces and with the general type of operators (including pseudodifferential)
in such bases.
Our problem as many related problems in the framework of
our type of approximations of complicated physical nonlinearities
is reduced to the problem of the solving of the systems of
differential equations with polynomial nonlinearities with or
without some constraints.
In this paper we consider as the main example the particle motion in
storage rings in standard approach, which is based on
consideration in [1], [2].
Starting from Hamiltonian, which described classical dynamics in
storage rings
\begin{eqnarray}
{\cal H}(\vec{r},\vec{P},t)=c\{\pi^2+m_0^2c^2\}^{1/2}+e\phi
\end{eqnarray}
and using Serret--Frenet parametrization, we have the following
Hamiltonian for orbital motion in machine coordinates:

\begin{eqnarray}
{\cal H}(x,p_x,&z&,p_z,\sigma,p_\sigma;s)=
   p_\sigma-[1+f(p_\sigma]\cdot[1+K_x\cdot x+K_z\cdot z]\times\\ \nonumber
& &\Bigg\{ 1-\frac{[p_x+H\cdot z]^2 + [p_z-H\cdot x]^2}
{[1+f(p_\sigma)]^2}\Bigg\}^{1/2}\\\nonumber
& &+\frac{1}{2}\cdot[1+K_x\cdot x+K_z\cdot z]^2-\frac{1}{2}\cdot g\cdot
   (z^2-x^2)-N\cdot xz\\ \nonumber
& &+\frac{\lambda}{6}\cdot(x^3-3xz^2)
   +\frac{\mu}{24}\cdot(z^4-6x^2z^2+x^4)\\ \nonumber
& &+\frac{1}{\beta_0^2}\cdot\frac{L}{2\pi\cdot h}\cdot\frac{eV(s)}{E_0}\cdot
\cos\left[h\cdot\frac{2\pi}{L}\cdot\sigma+\varphi\right]\nonumber
\end{eqnarray}

Then, after standard manipulations with truncation of
power series expansion of square root we arrive to the following
approximated Hamiltonian for particle motion:

\begin{eqnarray}
{\cal H}&=&
   \frac{1}{2}\cdot\frac{[p_x+H\cdot z]^2 + [p_z-H\cdot x]^2}
{[1+f(p_\sigma)]}+
p_\sigma-[1+K_x\cdot x+K_z\cdot z]\\\nonumber&&\cdot f(p_\sigma)+
\frac{1}{2}\cdot[K_x^2+g]\cdot x^2+\frac{1}{2}\cdot[K_z^2-g]\cdot z^2-
  N\cdot xz+\\ \nonumber
& &\frac{\lambda}{6}\cdot(x^3-3xz^2)+\frac{\mu}{24}\cdot(z^4-6x^2z^2+x^4)\\ \nonumber
& &+\frac{1}{\beta_0^2}\cdot\frac{L}{2\pi\cdot h}\cdot\frac{eV(s)}{E_0}\cdot
\cos\left[h\cdot\frac{2\pi}{L}\cdot\sigma+\varphi\right]\nonumber
\end{eqnarray}

and the corresponding equations of motion:

\begin{eqnarray}
\frac{d}{ds}x&=&\frac{\partial{\cal H}}{\partial p_x}=
     \frac{p_x+H\cdot z}{[1+f(p_\sigma)]}; \nonumber\\
\frac{d}{ds}p_x&=&-\frac{\partial{\cal H}}{\partial x}=
\frac{[p_z-H\cdot x]}{[1+f(p_\sigma)]}\cdot H -[K^2_x+g]\cdot x+N\cdot z+
 \nonumber\\
  & & K_x\cdot f(p_\sigma)-\frac{\lambda}{2}\cdot(x^2-z^2)-
  \frac{\mu}{6}(x^3-3xz^2);\\
\frac{d}{ds}z&=&\frac{\partial{\cal H}}{\partial p_z}=
 \frac{p_z-H\cdot x}{[1+f(p_\sigma)]}; \nonumber\\
\frac{d}{ds}p_z&=&-\frac{\partial{\cal H}}{\partial z}=
     -\frac{[p_x+H\cdot z]}{[1+f(p_\sigma)]}\cdot H -
     [K^2_z-g]\cdot z+N\cdot x+ \nonumber\\
  & & K_z\cdot f(p_\sigma)-\lambda\cdot xz-
  \frac{\mu}{6}(z^3-3x^2z);\nonumber\\
\frac{d}{ds}\sigma&=&\frac{\partial{\cal H}}{\partial p_\sigma}=
 1-[1+K_x\cdot x+K_z\cdot z]\cdot f^\prime(p_\sigma)- \nonumber\\
& &\frac{1}{2}\cdot\frac{[p_x+H\cdot z]^2+[p_z-H\cdot x]^2}{[1+f(p_\sigma)]^2}
 \cdot f^\prime(p_\sigma)\nonumber\\
\frac{d}{ds}p_\sigma&=&-\frac{\partial{\cal H}}{\partial\sigma}=
\frac{1}{\beta_0^2}\cdot\frac{eV(s)}{E_0}\cdot\sin\left[h\cdot\frac{2\pi}{L}
\cdot\sigma+\varphi\right] \nonumber
\end{eqnarray}

Then we use series expansion of function $f(p_\sigma)$ from [2]:
$$
f(p_\sigma)=f(0)+f^\prime(0)p_\sigma+f^{\prime\prime}(0)
\frac{1}{2}p_\sigma^2+\ldots=p_\sigma-\frac{1}{\gamma_0^2}
\cdot\frac{1}{2}p_\sigma^2+\ldots
$$
and the corresponding expansion of RHS of equations (4).
In the following we take into account only an arbitrary
polynomial (in terms of dynamical variables) expressions and
neglecting all nonpolynomial types of expressions, i.e. we
consider such approximations of RHS, which are not more than polynomial
functions in dynamical variables and arbitrary functions of
independent variable $s$ ("time" in our case, if we consider
our system of equations as dynamical problem).

\section{Polynomial Dynamics}
\subsection*{Introduction.}
The first main part of our consideration is some variational approach
to this problem, which reduces initial problem to the problem of
solution of functional equations at the first stage and some
algebraical problems at the second stage.
We consider also two private cases of our general construction.
In the first case (particular) we have for Riccati equations
(particular quadratic approximations) the solution
as a series on shifted
Legendre polynomials, which is parameterized by the solution
of reduced algebraical (also Riccati) system of equations.
This is only an example of general construction.
In the second case (general polynomial system) we have
the solution in a compactly
supported wavelet basis.
Multiresolution expansion is the second main part of our construction.
The solution is parameterized by solutions of two reduced algebraical
problems, one as in the first case and the second is some linear
problem, which is obtained from one of the next wavelet
construction: Fast Wavelet Transform (FWT), Stationary
Subdivision Schemes (SSS), the method of Connection
Coefficients (CC).

\subsection*{ Variational method.}
Our problems may be formulated as the systems of ordinary differential
equations
$
{dx_i}/{dt}=f_i(x_j,t), \quad (i,j=1,...,n)
$
with fixed initial conditions $x_i(0)$, where $f_i$ are not more
than polynomial functions of dynamical variables $x_j$
and  have arbitrary dependence of time. Because of time dilation
we can consider  only next time interval: $0\leq t\leq 1$.
 Let us consider a set of
functions
$
 \Phi_i(t)=x_i{dy_i}/{dt}+f_iy_i
$
and a set of functionals
$
F_i(x)=\int_0^1\Phi_i (t)dt-x_iy_i\mid^1_0,
$
where $y_i(t) (y_i(0)=0)$ are dual variables.
It is obvious that the initial system  and the system $F_i(x)=0$
are equivalent.
In the last part   we consider the symplectization of this approach.
Now we consider formal expansions for $x_i, y_i$:
\begin{eqnarray}
x_i(t)=x_i(0)+\sum_k\lambda_i^k\varphi_k(t)\quad
y_j(t)=\sum_r \eta_j^r\varphi_r(t),
\end{eqnarray}
where because of initial conditions we need only $\varphi_k(0)=0$.
Then we have the following reduced algebraical system
of equations on the set of unknown coefficients $\lambda_i^k$ of
expansions (5):
\begin{eqnarray}
\sum_k\mu_{kr}\lambda^k_i-\gamma_i^r(\lambda_j)=0
\end{eqnarray}
Its coefficients are
$
\mu_{kr}=\int_0^1\varphi_k'(t)\varphi_r(t)dt,\quad
\gamma_i^r=\int_0^1f_i(x_j,t)\varphi_r(t)dt.
$
Now, when we solve system (6) and determine
 unknown coefficients from formal expansion (5) we therefore
obtain the solution of our initial problem.
It should be noted if we consider only truncated expansion (5) with N terms
then we have from (6) the system of $N\times n$ algebraical equations and
the degree of this algebraical system coincides
 with degree of initial differential system.
So, we have the solution of the initial nonlinear
(polynomial) problem  in the form
\begin{eqnarray}
x_i(t)=x_i(0)+\sum_{k=1}^N\lambda_i^k X_k(t),
\end{eqnarray}
where coefficients $\lambda_i^k$ are roots of the corresponding
reduced algebraical problem (6).
Consequently, we have a parametrization of solution of initial problem
by solution of reduced algebraical problem (6). But in general case,
when the problem of computations of coefficients of reduced algebraical
system (6) is not  solved explicitly as in the quadratic case, which we
shall consider below, we have also parametrization of solution (4) by
solution of corresponding problems, which appear when we need to calculate
coefficients of (6).
As we shall see, these problems may be explicitly solved in wavelet approach.

\subsection*{The solutions}
Next we consider the  construction  of explicit time
solution for our problem. The obtained solutions are given
in the form (7),
where in our first case we have
$X_k(t)=Q_k(t)$, where $Q_k(t)$ are shifted Legendre
polynomials  and $\lambda_k^i$ are roots of reduced
quadratic system of equations.  In wavelet case $X_k(t)$
correspond to multiresolution expansions in the base of
 compactly supported wavelets and $\lambda_k^i$ are the roots of
corresponding general polynomial  system (6)  with coefficients, which
are given by FWT, SSS or CC constructions.  According to the
        variational method   to  give the reduction from
differential to algebraical system of equations we need compute
the objects $\gamma ^j_a$ and $\mu_{ji}$,
which are constructed from objects:
\begin{eqnarray}
\sigma_i&\equiv&\int^1_0X_i(\tau)d\tau=(-1)^{i+1},\quad
   \nu_{ij}\equiv\int^1_0X_i(\tau)X_j(\tau)d\tau=
    \sigma_i\sigma_j+\frac{\delta_{ij}}{ (2j+1)},\nonumber\\
    \mu_{ji}&\equiv&\int
X'_i(\tau)X_j(\tau)d\tau=\sigma_jF_1(i,0)+F_1(i,j),\\
    & &F_1(r,s)=[1-(-1)^{r+s}]\hat{s}(r-s-1),\quad
    \hat{s}(p)=\left\{ \begin{array}{ll}
       1, & \quad p\geq 0\\
       0, & \quad p < 0
       \end{array}
\right. \nonumber \\
   \beta_{klj}&\equiv&\int^1_0X_k(\tau)X_l(\tau)X_j(\tau)
   d\tau=\sigma_k\sigma_l\sigma_j+ \nonumber\\
& & \alpha_{klj}+
   \frac{\sigma_k\delta_{jl}}{ 2j+1}+
      \frac{\sigma_l\delta_{kj}}{ 2k+1}+
       \frac {\sigma_j\delta_{kl}}{ 2l+1}, \nonumber\\
      \alpha_{klj}&\equiv&\int^1_0X^\ast_kX^\ast_lX^\ast_jd\tau=
   \frac{1}{ (j+k+l+1)R\bigl(1/2(i+j+k)\bigr)}\times\nonumber\\
& &    R\bigl(1/2(j+k-l)\bigr)
    R\bigl(1/2(j-k+l)\bigr)
    R\bigl(1/2(-j+k+l)\bigr),\nonumber
\end{eqnarray}
if $j+k+l=2m , m\in{\it Z}$,and
  $\alpha_{klj}=0$ if $ j+k+l=2m+1 $;
where
$ R(i)=(2i)!/(2^ii!)^2$, $Q_i=\sigma_i+P_i^\ast$, where the
second equality in the formulae for $\sigma, \nu, \mu, \beta,\alpha$
hold for the first case.

\subsection*{ Wavelet computations.}
 Now we give construction for
computations of objects (8) in the wavelet case.
We use some constructions from multiresolution analysis: a sequence of
successive approximation closed subspaces $V_j$:
$
...V_2\subset V_1\subset V_0\subset V_{-1}\subset V_{-2}\subset ...
$
satisfying the following properties:
$
\displaystyle\bigcap_{j\in{\bf Z}}V_j=0$,
$\overline{\displaystyle\bigcup_{j\in{\bf Z}}}V_j=L^2({\bf R})$,
$ f(x)\in V_j <=> f(2x)\in V_{j+1}
$
There is a function $\varphi\in V_0$ such that \{${\varphi_{0,k}(x)=
\varphi(x-k)}_{k\in{\bf Z}}$\} forms a Riesz basis for $V_0$.
We use compactly supported wavelet basis: orthonormal basis
for functions in $L^2({\bf R})$. As usually $\varphi(x)$ is
a scaling function, $\psi(x)$ is a wavelet function, where
$\varphi_i(x)=\varphi(x-i)$. Scaling relation that defines
$\varphi,\psi$ are
\begin{eqnarray*}
\varphi(x)=\sum\limits^{N-1}_{k=0}a_k\varphi(2x-k)=
\sum\limits^{N-1}_{k=0}a_k\varphi_k(2x),\quad
\psi(x)=\sum\limits^{N-2}_{k=-1}(-1)^k a_{k+1}\varphi(2x+k)
\end{eqnarray*}
Let be  $ f : {\bf R}\longrightarrow {\bf C}$ and the wavelet
expansion is
\begin{eqnarray}
f(x)=\sum\limits_{\ell\in{\bf Z}}c_\ell\varphi_\ell(x)+
\sum\limits_{j=0}^\infty\sum\limits_{k\in{\bf
Z}}c_{jk}\psi_{jk}(x)
\end{eqnarray}
The indices $k, \ell$ and $j$ represent translation and scaling, respectively
$$
\varphi_{jl}(x)=2^{j/2}\varphi(2^j x-\ell),
\psi_{jk}(x)=2^{j/2}\psi(2^j x-k)
$$
The set $\{\varphi_{j,k}\}_{k\in {\bf Z}}$ forms a Riesz basis for $V_j$.
Let $W_j$ be the orthonormal complement of $V_j$ with respect to $V_{j+1}$.
Just as $V_j$ is spanned by dilation and translations of the scaling function,
so are $W_j$ spanned by translations and dilation of the mother wavelet
$\psi_{jk}(x)$.
If in formulae (9) $c_{jk}=0$ for $j\geq J$, then $f(x)$ has an alternative
expansion in terms of dilated scaling functions only
$
f(x)=\sum\limits_{\ell\in {\bf Z}}c_{J\ell}\varphi_{J\ell}(x)
$.
This is a finite wavelet expansion, it can be written solely in
terms of translated scaling functions. We use wavelet $\psi(x)
$, which has $k$ vanishing moments
$
\int x^k \psi(x)d(x)=0$, or equivalently
$x^k=\sum c_\ell\varphi_\ell(x)$ for each $k$,
$0\leq k\leq K$.
Also we have the shortest possible support: scaling function
$DN$ (where $N$ is even integer) will have support $[0,N-1]$ and
$N/2$ vanishing moments.
There exists $\lambda>0$ such that $DN$ has $\lambda N$
continuous derivatives; for small $N,\lambda\geq 0.55$.
To solve our second associated linear problem we need to
evaluate derivatives of $f(x)$ in terms of $\varphi(x)$.
Let be $
\varphi^n_\ell=d^n\varphi_\ell(x)/dx^n
$.
We derive the wavelet - Galerkin approximation of a
differentiated $f(x)$ as
$
f^d(x)=\sum_\ell c_l\varphi_\ell^d(x)
$
and values $\varphi_\ell^d(x)$ can be expanded in terms of
$\varphi(x)$
\begin{eqnarray*}
\phi_\ell^d(x)=\sum\limits_m\lambda_m\varphi_m(x),\quad
\lambda_m=\int\limits_{-\infty}^{\infty}\varphi_\ell^d(x)\varphi_m(x)dx
 \end{eqnarray*}
The coefficients $\lambda_m$  are 2-term connection
coefficients. In general we need to find $(d_i\geq 0)$
\begin{eqnarray}
\Lambda^{d_1 d_2 ...d_n}_{\ell_1 \ell_2 ...\ell_n}=
 \int\limits_{-\infty}^{\infty}\prod\varphi^{d_i}_{\ell_i}(x)dx
\end{eqnarray}
For Riccati case we need to evaluate two and three
connection coefficients
\begin{eqnarray*}
\Lambda_\ell^{d_1
d_2}=\int^\infty_{-\infty}\varphi^{d_1}(x)\varphi_\ell^{d_2}(x)dx,
\quad
\Lambda^{d_1 d_2
d_3}=\int\limits_{-\infty}^\infty\varphi^{d_1}(x)\varphi_
\ell^{d_2}(x)\varphi^{d_3}_m(x)dx
\end{eqnarray*}
According to CC method [12] we use the next construction. When $N$  in
scaling equation is a finite even positive integer the function
$\varphi(x)$  has compact support contained in $[0,N-1]$.
For a fixed triple $(d_1,d_2,d_3)$ only some  $\Lambda_{\ell
 m}^{d_1 d_2 d_3}$ are nonzero : $2-N\leq \ell\leq N-2,\quad
2-N\leq m\leq N-2,\quad |\ell-m|\leq N-2$. There are
$M=3N^2-9N+7$ such pairs $(\ell,m)$. Let $\Lambda^{d_1 d_2 d_3}$
be an M-vector, whose components are numbers $\Lambda^{d_1 d_2
d_3}_{\ell m}$. Then we have the first key result: $\Lambda$
satisfy the system of equations $(d=d_1+d_2+d_3)$
\begin{eqnarray*}
A\Lambda^{d_1 d_2 d_3}=2^{1-d}\Lambda^{d_1 d_2 d_3},
\qquad
A_{\ell,m;q,r}=\sum\limits_p a_p a_{q-2\ell+p}a_{r-2m+p}
\end{eqnarray*}
By moment equations we have created a system of $M+d+1$
equations in $M$ unknowns. It has rank $M$ and we can obtain
unique solution by combination of LU decomposition and QR
algorithm.
The second key result gives us the 2-term connection
coefficients:
\begin{eqnarray*}
A\Lambda^{d_1 d_2}=2^{1-d}\Lambda^{d_1 d_2},\quad d=d_1+d_2,\quad
A_{\ell,q}=\sum\limits_p a_p a_{q-2\ell+p}
\end{eqnarray*}
For nonquadratic case we have analogously additional linear problems for
objects (10).
Also, we use FWT and SSS for computing coefficients of reduced
algebraic systems.
We use for modelling D6,D8,D10 functions and programs RADAU and
DOPRI for testing.

As a result we obtained the explicit time solution (7) of our
 problem. In comparison with
 wavelet expansion
on the real line which we use now and in
calculation of Galerkin approximation, Melnikov function approach, etc also
we need to use periodized wavelet expansion, i.e. wavelet expansion
 on finite interval. Also in the solution of perturbed system we have
some problem with variable coefficients.
For solving last problem we need to consider one more refinement equation
for scaling function $\phi_2(x)$:
$
\phi_2 (x)=\sum\limits^{N-1}_{k=0} a^2_k \phi_2 (2x-k)
$
and corresponding wavelet expansion for variable coefficients $b(t)$:
$
\sum\limits_k B_k^j (b)\phi_2(2^jx-k),
$
where $B_k^j (b)$ are functionals supported in a small neighborhood of
$2^{-j}k$.

The solution of the first problem consists in periodizing. In this case
we use expansion  into periodized wavelets
defined by
$
\phi^{per}_{-j,k}(x)=2^{j/2}\sum\limits_{Z} \phi(2^jx+2^j\ell-k).
$
All these modifications lead only to transformations of coefficients of
reduced algebraic system, but general scheme remains the same.

\section{Metaplectic Wavelets}
In this part we continue the application of powerful methods of wavelet
analysis to polynomial approximations of nonlinear accelerator 
physics problems. In  part 1 we considered our main example
and general approach
for constructing wavelet representation for orbital motion in
storage rings. But now we need take into account the Hamiltonian
or symplectic structure related with system (4).
Therefore, we need to consider generalized wavelets, which
allow us to consider the corresponding symplectic structures, instead of
compactly supported wavelet representation.
By using the orbit method and  constructions
from the geometric quantization theory we consider the
symplectic  and Poisson structures associated with Weyl--\-
 Heisenberg wavelets by using metaplectic structure and
the corresponding polarization. In the next part  we consider  applications
 to construction of Melnikov functions in the theory of
 homoclinic chaos in perturbed Hamiltonian systems.

In wavelet analysis the following three concepts are used now:
1).\ a square integrable representation $U$ of a group $G$, 
2).\ coherent states over G,  
3).\ the wavelet transform associated to U.

We have three important particular cases:\\
a) the affine $(ax+b)$ group, which yields the usual wavelet
analysis
$$
[\pi(b,a)f](x)=\frac{1}{\sqrt{a}}f\left(\frac{x-b}{a}\right)
$$
b). the Weyl-Heisenberg  group which leads to the Gabor
functions, i.e. coherent states associated with windowed Fourier
transform.
$$
[\pi(q,p,\varphi)f](x)=\exp(i\mu(\varphi-p(x-q))f(x-q)
$$
In both cases time-frequency plane corresponds to the phase
space of group representation.\\
c). also, we have
 the case of bigger group, containing
both affine and We\-yl-\-Hei\-sen\-berg group, which interpolate between
affine wavelet analysis and windowed Fourier analysis: affine
Weyl--Heisenberg group [13]. But usual representation of it is not
square--integrable and must be modified: restriction of the
representation to a suitable quotient space of the group (the
associated phase space in that case) restores square --
integrability:
 $G_{aWH\longrightarrow}$ homogeneous space.
Also, we have more general approach which allows to consider wavelets 
corresponding to more general groups and representations [14], [15].
Our goal is applications of these results to problems of 
Hamiltonian dynamics and as consequence we need to take into account 
symplectic nature of our dynamical problem. 
 Also, the symplectic and wavelet structures
 must be consistent (this must
be resemble the symplectic or Lie-Poisson integrator theory).
 We use the
point of view of geometric quantization theory (orbit method)
instead of harmonic analysis. Because of this we can consider
(a) -- (c) analogously.

\subsection*{Metaplectic Group and Representations.}
Let $Sp(n)$ be
symplectic group, $Mp(n)$ be its unique two- fold covering --
metaplectic group.   Let V be a symplectic vector space
with symplectic form ( , ), then $R\oplus V$ is nilpotent Lie
algebra - Heisenberg algebra:
$$[R,V]=0, \quad [v,w]=(v,w)\in
R,\quad [V,V]=R.$$
$Sp(V)$ is a group of automorphisms of
Heisenberg algebra.

 Let N be a group with Lie algebra $R\oplus
V$, i.e.  Heisenberg group.  By Stone-- von Neumann theorem
Heisenberg group has unique irreducible unitary representation
in which $1\mapsto i$. This representation is projective:
$U_{g_1}U_{g_2}=c(g_1,g_2)\cdot U_{g_1g_2}$, where c is a map:
$Sp(V)\times Sp(V)\rightarrow S^1$, i.e. c  is $S^1$-cocycle.

But this representation is unitary representation of universal
covering, i.e. metaplectic group $Mp(V)$. We give this
representation without Stone-von Neumann theorem.\
Consider a new group $F=N'\bowtie Mp(V),\quad \bowtie$ is semidirect
product (we consider instead of  $ N=R\oplus V$ the $
N'=S^1\times V, \quad S^1=(R/2\pi Z)$). Let $V^*$ be dual to V,
$G(V^*)$ be automorphism group of $V^*$.Then F is subgroup of $
G(V^*)$, which consists of elements, which acts on $V^*$ by affine
transformations. \\
This is the key point!

 Let $q_1,...,q_n;p_1,...,p_n$ be symplectic basis in V,
$\alpha=pdq=\sum p_{i}dq_i $  and $d\alpha$ be symplectic form on
$V^*$. Let M be fixed affine polarization, then for $a\in F$ the
map $a\mapsto \Theta_a$ gives unitary representation of G:
$ \Theta_a: H(M) \rightarrow H(M) $

Explicitly  we have for representation of N on H(M):
$$
(\Theta_qf)^*(x)=e^{-iqx}f(x),  \quad
 \Theta_{p}f(x)=f(x-p)
$$
The representation of N on H(M) is irreducible. Let $A_q,A_p$
be infinitesimal operators of this representation
$$
 A_q=\lim_{t\rightarrow 0} \frac{1}{t}[\Theta_{-tq}-I], \quad
 A_p=\lim_{t\rightarrow 0} \frac{1}{t}[\Theta_{-tp}-I],
$$
$$\mbox{then}\qquad
A_q f(x)=i(qx)f(x),\quad A_p f(x)=\sum p_j\frac{\partial
f}{\partial x_j}(x)
$$
Now we give the representation of infinitesimal ba\-sic
elements. Lie algebra of the group F is the algebra of all
(non\-ho\-mo\-ge\-ne\-ous) quadratic po\-ly\-no\-mi\-als of (p,q) relatively
Poisson bracket (PB). The basis of this algebra consists of
elements
$1,q_1,...,q_n$,\  $p_1,...,p_n$,\ $ q_i q_j, q_i p_j$,\ $p_i p_j,
 \quad i,j=1,...,n,\quad i\leq j$,
 \begin{eqnarray*}
 & &PB \ is
\quad \{ f,g\}=\sum\frac{\partial f}{\partial p_j}
\frac{\partial g}{\partial q_i}-\frac{\partial f}{\partial q_i}
\frac{\partial g}{\partial p_i} \quad
 \mbox{and}  \quad
   \{1,g \}= 0 \quad for \mbox{ all} \ g,\\
& &\{ p_i,q_j\}= \delta_{ij},\quad \{p_i
q_j,q_k\}=\delta_{ik}q_j,\quad
    \{p_i q_j,p_k\}=-\delta_{jk}p_i, \quad \{p_ip_j,p_k\}=0,\\
& & \{p_i p_j,q_k \}=
\delta_{ik}p_j+\delta_{jk}p_i,\quad
  \{ q_i q_j,q_k\}=0,
 \{q_i q_j,p_k\}=-\delta_{ik}q_j-\delta_{jk}q_i
 \end{eqnarray*}
so, we have the representation of basic elements
 $ f\mapsto A_f : 1\mapsto i, q_k\mapsto ix_k $,
\begin{eqnarray*}
 p_l\mapsto\frac{\delta}{\delta x^l}, p_i q_j\mapsto
x^i\frac{\partial}{\partial x^j}+\frac{1}{2}\delta_{ij},\qquad
 p_k p_l\mapsto \frac{1}{i}\frac{\partial^k}{\partial x^k\partial
x^l}, q_k q_l\mapsto ix^k x^l
\end{eqnarray*}
This gives  the structure of the Poisson mani\-folds to
representation of any (nilpotent) algebra or in other words to
continuous wavelet trans\-form.

\subsection*{ The Segal-Bargman Representation.}
Let $ z=1/\sqrt{2}\cdot(p-iq),\quad
\bar{z}=1/\sqrt{2}\cdot(p+iq),\quad
$
$ p=(p_1,...,p_n)
,\quad
 F_n $ is the
space of holomorphic functions of n complex variables with
$(f,f)< \infty$, where $$ (f,g)=(2\pi)^{-n}\int
f(z)\overline{g(z)}e^{-|z|^2}dpdq $$
Consider a map  $U:
H\rightarrow F_n$ , where H is with real polarization, $F_n
$ is with complex polarization, then we have $$(U\Psi)(a)=\int
A(a,q)\Psi(q)dq,\qquad \mbox{where}\quad
A(a,q)=\displaystyle\pi^{-n/4}e^{-1/2(a^2+q^2)+\sqrt{2}aq}
$$
i.e. the Bargmann formula produce  wavelets.We also have
the representation of Heisenberg algebra on $F_n$ :
\begin{eqnarray*}
U\frac{\partial}{\partial q_j} U^{-1}=\frac{1}{\sqrt{2}}\left
(z_j- \frac{\partial}{\partial z_j}\right),\qquad
 Uq_j
U^{-1}=-\frac{i}{\sqrt{2}}\left(z_j+\frac{\partial }{\partial
 z_{j}} \right)
\end{eqnarray*}
and also : $ \omega=d\beta=dp\wedge dq,$
 where
$\beta=i\bar{z}dz $.

\subsection*{ Orbital Theory for  Wavelets.}
Let coadjoint action be
$<g\cdot f,Y>=<f,Ad(g)^{-1}Y>,$
 where $<,>$ is pairing
$ g\in G,\quad f\in g^*,\quad Y\in{\cal G}$.
The orbit is
${\cal O}_f=G\cdot f\equiv G/G(f)$.
Also, let A=A(M) be algebra of functions,
V(M) is A-module of vector fields,
$A^p$  is A-module of p-forms.
Vector fields on orbit is
$$
\sigma({\cal O},X)_f(\phi)=\frac{d}{dt}(\phi(\exp tXf))\Big |_{t=0}
$$
where $\phi\in A({\cal O}),\quad f\in{\cal O}$. Then ${\cal O}_f$
are homogeneous symplectic manifolds with 2-form
$
\Omega(\sigma({\cal O},X)_f,\sigma({\cal  O},Y)_f)=<f,[X,Y]>,
$
and $d\Omega=0$. PB on ${\cal O}$ have the next form
$
\{ \Psi_1,\Psi_2\}=p(\Psi_1)\Psi_2
$
where p is $ A^1({\cal O})\rightarrow V({\cal O})$ with
definition
$\Omega (p(\alpha),X)$ $=$ $i(X)\alpha$. Here $\Psi_1,\Psi_2\in
A(\cal {O})$ and  $A({\cal O}) $ is Lie algebra with bracket
\{,\}.
Now let N be a Heisenberg group. Consider adjoint and
coadjoint representations in some particular case.
 $N=(z,t)\in C\times R,
 z=p+iq$; compositions in N are $(z,t)\cdot(z',t')=
 (z+z',t+t'+B(z,z')) $, where $B(z,z')=pq'-qp'$. Inverse
 element is $(-t,-z)$. Lie algebra n of N is  $(\zeta,\tau)
 \in C\times R$ with bracket $[(\zeta,\tau),(\zeta',\tau')]=
 (0,B(\zeta,\zeta'))$. Centre is $\tilde{z}\in n $ and
generated by (0,1);
 Z is a subgroup $\exp\tilde{z}$.
Adjoint representation N on n is given by formula
  $
Ad(z,t)(\zeta,\tau)=(\zeta,\tau+B(z,\zeta))
$
Coadjoint:
 for $f\in n^*,\quad g=(z,t)$,
$(g \cdot f)(\zeta,\zeta)=f(\zeta,\tau)-B(z,\zeta)f(0,1)$ then
  orbits for which $f|_{\tilde z}\neq 0$ are plane in $n^*$
 given by equation $ f(0,1)=\mu$ . If $X=(\zeta,0),\quad
 Y=(\zeta ',0),\quad X,Y\in n$ then symplectic structure
  is
\begin{eqnarray*}
  \Omega (\sigma({\cal O},X)_f,\sigma({\cal
  O},Y)_f)=<f,[X,Y]>=
  f(0,B(\zeta,\zeta'))\mu B(\zeta,\zeta')
\end{eqnarray*}
Also we have for orbit ${\cal O}_\mu=N/Z$ and ${\cal O}_\mu $ is
Hamiltonian G-space.

\subsection*{ Kirillov Character Formula  or
 Analogy of Gabor Wavelets.}
Let U denote irreducible unitary representation of N with
condition $U(0,t)=\exp(it\ell)\cdot 1$, where $ \ell\neq
0 $,then U is equivalent to representation $T_\ell$ which acts in
$L^2(R)$ according to
$$
T_\ell(z,t)\phi(x)=\exp\left(i\ell(t+px)\right)\phi(x-q)
$$
If instead of N we consider E(2)/R we have $S^1$ case and we
  have Gabor functions on $S^1$.

\subsection*{ Oscillator Group.}
Let O be an oscillator group,i.e. semidirect product of R
and Heisenberg group N.
Let H,P,Q,I be standard basis in Lie algebra o of the group O
and $H^*,P^*,Q^*,I^*$ be dual basis in $o^*$. Let functional
f=(a,b,c,d)  be
$
aI^*+bP^*+cQ^*+dH^* .
$
Let us consider complex polarizations
$
h=( H,I,P+iQ ),     \quad 
\bar{h}=(I,H,P-iQ)
$
Induced from $h$ representation, corresponding to functional f
(for $a>0$), unitary equivalent to the representation
$$
W(t,n)f(y)=\exp (it(h-1/2)) \cdot U_{a} (n)V(t),
$$
\begin{eqnarray*}
\mbox{where}\qquad V(t)=\exp[-it(P^2+Q^2)/2a] ,\quad
 P=-{d}/{dx} ,\quad
 Q=iax,
\end{eqnarray*}
and $U_a(n)$ is irreducible representation of N, which have the
form $U_a(z)=exp(iaz)$ on the center of N.
Here we have:
U(n=(x,y,z)) is Schr\"{o}\-din\-ger representation, $U_t(n)=U(t(n))$
is the representation obtained from previous by
automorphism (time translation)
$n\longrightarrow t(n);\quad U_t(n)=U(t(n))$
is also unitary irreducible representation of N.
$
V(t)=\exp(it(P^2+Q^2+h-1/2))
$
is an operator, which according to Stone--von Neumann theorem
has the property
$
U_t(n)=V(t)U(n)V(t)^{-1}.
$

This is our last private case, but according to our approach we
can construct by using methods of geometric quantization theory
many "symplectic wavelet constructions" with corresponding
symplectic or Poisson structure on it.
Very useful particular spline--wavelet basis with uniform
exponential control on stratified and nilpotent Lie groups
was considered in [15].

\section{Melnikov Functions Approach}
In this part we continue the application of the methods of wavelet
analysis to polynomial approximations of nolinear accelerator
physics problems.
Now we consider one problem of nontrivial dynamics
related with complicated differential geometrical and  topological
structures of
system (4).  We give some points of applications
of wavelet methods from the preceding parts to Melnikov approach
in the theory of
homoclinic chaos in perturbed Hamiltonian systems.

\subsection*{ Routes to Chaos}
  Now we give some points of our program of
 understanding routes to  chaos in some Hamiltonian
systems in the wavelet approach [3]-[11]. All points are:
\begin{enumerate}
 \item A model.
 \item A computer zoo. The understanding of
the  computer zoo.
 \item  A naive Melnikov function approach.
 \item A naive wavelet description of (hetero) homoclinic orbits
(separatrix) and quasiperiodic oscillations.
 \item Symplectic Melnikov function approach.
 \item Splitting of separatrix... $\longrightarrow$stochastic web
 with magic symmetry, Arnold diffusion and all
that.
\end{enumerate}
1. As a model we have two frequencies perturbations of particular
case of system (4):
\begin{eqnarray*}
\dot x_1&=&x_2\qquad \dot x_3=x_4,\qquad \dot x_5=1,\qquad \dot x_6=1,\nonumber\\
\dot x_2&=&-ax_1-b[\cos(rx_5)+\cos(sx_6)]x_1-dx^3_1-
    mdx_1x^2_3-
    px_2-\varphi(x_5)  \nonumber\\
\dot x_4&=&ex_3-f[\cos(rx_5)+\cos(sx_6)]x_3- gx_3^3-
  kx_1^2x_3-gx_4-
   \psi(x_5) \nonumber 
\end{eqnarray*}
or in  Hamiltonian form
\begin{eqnarray*}
\dot{x}=J\cdot\nabla H(x)+\varepsilon g(x,\Theta), \quad
\dot{\Theta}&=&\omega,\quad
(x,\Theta)\in R^4\times T^2,\quad
T^2=S^1\times S^1,
\end{eqnarray*}
for $\varepsilon=0 $ we have:
\begin{equation}
\dot{x}=J\cdot\nabla H(x),\quad
\dot\Theta=\omega
\end{equation}
2. For pictures and details one can see [5], [10].
The key point is the
splitting of separatrix (homoclinic orbit) and transition to
fractal sets on the Poincare sections.\\
3. For $\varepsilon=0$ we
have homoclinic orbit $\bar{x}_{0}(t)$ to the hyperbolic fixed
point $x_0$. For $\varepsilon\neq 0$ we have normally hyperbolic
invariant torus $T_{\varepsilon}$ and condition on transversally
intersection of stable and unstable ma\-ni\-folds
$W^s(T_{\varepsilon})$ and $W^u(T_{\varepsilon})$ in terms of
Melnikov functions $M(\Theta)$ for $\bar{x}_{0}(t)$.  $$
M(\Theta)=\displaystyle\int\limits_{-\infty}^{\infty}\nabla
H(\bar{x}_{0}(t)) \wedge g(\bar{x}_{0}(t),\omega t+\Theta)dt
$$
This condition has the next form:
\begin{eqnarray*}
M(\Theta_0)=0, \qquad
\sum\limits_{j=1}^{2}\omega_j\frac{\partial}{\partial\Theta_j}
M(\Theta_0)\neq0  
\end{eqnarray*}
According to the approach of Birkhoff-Smale-Wiggins  we
determined the region in parameter space in which we observe the
chaotic behaviour [5], [10].\\
4. If we cannot solve equations (11)
explicitly in time, then we use the wavelet approach from part 1
for the computations of  homoclinic (heteroclinic) loops as
the wavelet solutions of system (11).
For computations of quasiperiodic Melnikov functions
$$
M^{m/n}(t_0)=\int^{mT}_0 DH(x_\alpha(t))\wedge g(x_\alpha(t),t+t_0)dt
$$
  we used periodization of wavelet solution from part 1.\\
5. We also used symplectic Melnikov function approach
\begin{eqnarray*}
M_i(z)&=&\displaystyle\lim_{j\rightarrow\infty}\int\limits_{-T_j^*}
^{T_j}\{h_i,\hat{h}\}_{\Psi (t,z)}dt  \\
d_i(z,\varepsilon)&=&h_i(z^u_\varepsilon)-h_i(z^s_\varepsilon)=
\varepsilon M_i(z)+O(\varepsilon^2)
\end{eqnarray*}
where $\{,\}$ is the Poisson bracket,
$d_i(z,\varepsilon)$ is the Melnikov distance. So, we need symplectic
invariant wavelet expressions for Poisson brackets. The computations
 are produced  according to part 2.\\
6. Some hypothesis about
 strange symmetry of stochastic web
in multi-degree-of freedom Hamiltonian systems [11].

\section{Symplectic Topology and Wavelets}
Now we consider another type of wavelet approach which gives us
a possibility to parametrize Arnold--Weinstein curves or
closed loops in Hamiltonian systems by generalized refinement
equations or Quadratic Mirror Filters equations.

\subsection*{ Wavelet Parametrization in Floer Approach.}
Now we consider the generalization of our wavelet variational
approach  to the symplectic invariant calculation of
closed loops in Hamiltonian systems
[16]. We also have the parametrization of our solution by some
reduced algebraical problem but in contrast to the general case where
the solution is parametrized by construction based on scalar
refinement equation, in symplectic case we have
parametrization of the solution
by matrix problems -- Quadratic Mirror Filters equations [17].

The action functional for loops in the phase space is [16]
$$
F(\gamma)=\displaystyle\int_\gamma pdq-\int_0^1H(t,\gamma(t))dt
$$
The critical points of $F$ are those loops $\gamma$, which solve
the Hamiltonian equations associated with the Hamiltonian $H$
and hence are periodic orbits. By the way, all critical points of $F$ are
the saddle points of infinite Morse index, but surprisingly this approach  is
very effective. This will be demonstrated using several
variational techniques starting from minimax due to Rabinowitz
and ending with Floer homology. So, $(M,\omega)$ is symplectic
manifolds, $H: M \to R $, $H$ is Hamiltonian, $X_H$ is
unique Hamiltonian vector field defined  by
$$
\omega(X_H(x),\upsilon)=-dH(x)(\upsilon),\quad \upsilon\in T_xM,
\quad x\in M,
$$
where $ \omega$ is the symplectic structure.
A T-periodic solution $x(t)$ of the Hamiltonian equations
$$
\dot x=X_H(x) \quad \mbox{ on $M$}
$$
is a solution, satisfying the boundary conditions $x(T)$ $=x(0), T>0$.
Let us consider the loop space $\Omega=C^\infty(S^1, R^{2n})$,
where $S^1=R/{\bf Z}$, of smooth loops in $R^{2n}$.
Let us define a function $\Phi: \Omega\to R $ by setting
$$
\Phi(x)=\displaystyle\int_0^1\frac{1}{2}<-J\dot x, x>dt-
\int_0^1 H(x(t))dt, \quad x\in\Omega
$$
The critical points of $\Phi$ are the periodic solutions of $\dot x=X_H(x)$.
Computing the derivative at $x\in\Omega$ in the direction of $y\in\Omega$,
we find
\begin{eqnarray*}
\Phi'(x)(y)=\frac{d}{d\epsilon}\Phi(x+\epsilon y)\vert_{\epsilon=0}
=
\displaystyle\int_0^1<-J\dot x-\bigtriangledown H(x),y>dt
\end{eqnarray*}
Consequently, $\Phi'(x)(y)=0$ for all $y\in\Omega$ iff the loop $x$ satisfies
the equation
$$
-J\dot x(t)-\bigtriangledown H(x(t))=0,
$$
i.e. $x(t)$ is a solution of the Hamiltonian equations, which also satisfies
$x(0)=x(1)$, i.e. periodic of period 1. Periodic loops may be represented by
their Fourier series:
$$
x(t)=\displaystyle\sum_{k\in{\bf Z}}e^{k2\pi Jt}x_k, \quad x_k\in R^{2k},
$$
where $J$ is quasicomplex structure. We give relations between
quasicomplex structure and wavelets in [11].
But now we use  the  construction [17]
for loop parametrization. It is based on the theorem about
explicit bijection between the Quadratic Mirror Filters (QMF) and
the whole loop group: $LG: S^1\to G$.
In particular case we have relation  between {\bf QMF}-systems and
measurable functions
$\chi: S^1 \to U(2)$ satisfying

\begin{displaymath}
\chi(\omega+\pi)=\chi(\omega)\left [ \begin{array}{ll}
                                 0 & 1\\
                                 1 & 0
                              \end{array}\right ],
\end{displaymath}
in the next explicit form
\begin{eqnarray*}
\left [ \begin{array}{ll}
          \hat\Phi_0(\omega) & \hat\Phi_0(\omega+\pi)\\
          \hat\Phi_1(\omega) & \hat\Phi_1(\omega+\pi)
         \end{array}\right ]
      &=& \chi(\omega)\left [\begin{array}{ll}
             0 & 1\\
             1 & 0
         \end{array}\right ]
       +\chi(\omega+\pi)\left [\begin{array}{ll}
             0 & 0\\
             0 & 1
           \end{array}\right ],
\end{eqnarray*}
where
$$
\left |\hat\Phi_i(\omega)\right |^2+\left |\hat\Phi_i(\omega+\pi)\right |^2=2,
\quad i=0,1.
$$

Also, we have symplectic structure
on $LG$
$$
\omega(\xi,\eta)=\frac{1}{2\pi}\int_0^{2\pi}<\xi(\theta),\eta'(\theta)>d\theta
$$
So, we have the parametrization of periodic orbits (Arnold--Weinstein curves)
 by reduced QMF equations.

Extended version and related results may be found in [3]-[11].

One of us (M.G.Z.) would like to thank A.~Dragt, J.~Irwin, 
F.~Schmidt for discussions, Zohreh Parsa for many discussions and 
continued encouragement during and after workshop "New Ideas for 
Particle Accelerators" and Institute for Theoretical Physics,
University of California, Santa Barbara for hospitality.

This research was supported in part under "New Ideas for Particle 
Accelerators Program" NSF-
Grant No.~PHY94-07194.

 \end{document}